\documentclass[10pt]{article}

\usepackage[english]{babel}
\usepackage[a4paper,top=2cm,bottom=2cm,left=2cm,right=2cm,marginparwidth=1.75cm]{geometry}

\usepackage{amsmath}
\usepackage{graphicx}
\usepackage[colorlinks=true, allcolors=black]{hyperref}
\usepackage{booktabs} %professional tables
\usepackage{makecell} %for two lines in tables
\usepackage{amsfonts} %for math bbb
\usepackage{multirow}
\usepackage{subcaption}
\usepackage{soul}
\usepackage[export]{adjustbox}

\usepackage[dvipsnames]{xcolor} % for colours
\usepackage{overpic} %for labelling networks

\usepackage[title]{appendix}

\title{Skill and spatial mismatches for sustainable development in Brazil
}

\author{Anna K. Berryman$^{1,2}$, Joris B{\"u}cker$^{2,3}$, Fernanda Senra de Moura$^{2,4,5}$,\\ Pete Barbrook-Johnson$^{2,3,4}$, Marek Hanusch$^{5, *}$, Penny Mealy$^{2,3,5,6,*}$, \\ J. Doyne Farmer$^{2,3,6,7}$, and R. Maria del Rio-Chanona$^{8,9,10,\dagger}$}
\date{%
    $^1$Mathematical Institute, University of Oxford %
    $^2$Institute for New Economic Thinking at the Oxford Martin School, University of Oxford 
    $^3$Smith School of Enterprise and the Environment, University of Oxford %
    $^4$Environmental Change Institute, University of Oxford %
    $^5$World Bank Group, Washington DC %
    $^6$Santa Fe Institute, Santa Fe %
    $^7$Macrocosm %
    $^8$Complexity Science Hub, Vienna %
    $^9$The Bennett Institute for Public Policy, University of Cambridge %
    $^{10}$Dept. of Computer Science, University College London \\
    [2ex]%
    \today}

\begin{document}
\maketitle

\renewcommand{\thefootnote}{*} % Temporarily redefine to dagger
\footnotetext{The findings, interpretations, and conclusions expressed in this article do not necessarily reflect the views of the World Bank, the Executive Directors of the World Bank or the governments they represent.}
\renewcommand{\thefootnote}{\arabic{footnote}} % Reset to Arabic numerals
\renewcommand{\thefootnote}{$\dagger$} % Temporarily redefine to dagger
\footnotetext{correspondence: m.delriochanona@ucl.ac.uk}
\renewcommand{\thefootnote}{\arabic{footnote}} % Reset to Arabic numerals

\begin{abstract}
Structural change is necessary for all countries transitioning to a more environmentally sustainable economy, but what are the likely impacts on workers? Studies often find that green transition scenarios result in net positive job creation numbers overall but rarely provide insights into the more granular dynamics of the labour market. This paper combines a dynamic labour market simulation model with development scenarios focused on agriculture and green manufacturing. We study how, within the context of a green transition, productivity shifts in different sectors and regions, with differing environmental impacts, may affect and be constrained by the labour market in Brazil. By accounting for labour market frictions associated with skill and spatial mismatches, we find that productivity shocks, if not well managed, can exacerbate inequality. Agricultural workers tend to be the most negatively affected as they are less occupationally and geographically mobile. Our results highlight the importance of well-targeted labour market policies to ensure the green transition is just and equitable.

\end{abstract}

\section{Introduction}
%1.
%% Motivation 
All countries need to undertake some degree of structural change in order to transition to more environmentally sustainable economies. This will impact the labour market in different ways. While current studies suggest that more jobs are likely to be created than destroyed, newly created jobs may require different skills~\cite{Saussay2022} or arise in different locations~\cite{Lim2023, xie_2023_labour_dist_us} from those that disappear. Such skill or spatial-related mismatches can drive unemployment in some occupations and regions, while leaving unfilled job vacancies in others. Unmanaged green transitions can lead to declining income and greater inequality and deprivation~\cite{Neffke2022, Westin1989}, giving rise to `green discontent' in neglected regions~\cite{rodriguez2023green}. Consequently, examining the labour implications of environmental policy at occupational and regional levels is paramount. 

In this paper, we present a labour market model which incorporates labour market frictions related to skill and spatial mismatches. We apply the model to two distinct development pathways with different environmental outcomes and study their impact on unemployment and vacancies across occupations and regions in Brazil. While Brazil's energy matrix is one of the least carbon-intensive in the world, its emissions related to deforestation and agriculture make it one of the largest greenhouse gas emitters~\cite{UNEP2021}. Brazil's transition to an environmentally sustainable economy, and achieving their 2050 goal of carbon-neutrality~\cite{ndc_adjust2023}, depends on shifting away from deforestation-intensive activities~\cite{Hanusch2023}, as the Amazon Rainforest is one of the world's largest carbon sinks~\cite{trancoso2021changing}.

%2.
%% PREVIOUS RESEARCH 
Previous research has shown that shifts towards more environmentally sustainable development pathways will create and destroy jobs. Garrett-Peltier~\cite{GP2017} uses an input-output (IO) model and finds that every \$1 million shifted from brown to green energy destroys 2.65 brown jobs and creates 7.49 green jobs, causing a net increase of five jobs. Similarly, Montt et al.~\cite{ILO2018} find that a shift to net zero will cause net job creation. Analysis using computable general equilibrium (CGE) models to understand the net-zero transition find different aggregate labour market impacts. For example, Fragkos and Paroussos~\cite{Fragkos2018} find a net creation of 200,000 jobs in the energy sector, and Castellanos and Heutel~\cite{CastellanosHeutel2019} and Hafstead and Williams~\cite{Hafstead2018} find that, although the aggregate impacts are small, large shifts of labour between sectors will be required. 

However, these models do not account for labour frictions that workers encounter when they switch jobs during these large labour shifts. It has been shown that, considering labour frictions, the cost of climate policies might be higher than previously thought~\cite{Guivarch2011}, and these labour market frictions can be substantial, preventing workers from accessing jobs created by such policies~\cite{Donovan2021, Moyen2005, Vandeplas2022}. Additionally, analysis of job vacancies and skill descriptions suggests that disparities in skills and pay may occur during the green transition~\cite{sato2023skills, Saussay2022}. The mismatch in skills may negatively impact the wages of displaced workers~\cite{Neffke2022}, create `hard-to-fill' vacancies in specific occupations, and slow down transition pathways~\cite{Attstrom2014, bucker2023employment, Hakansson2020, Lankhuizen2023}. 

In addition to skill frictions, geographical frictions also influence levels of unemployment~\cite{Bilal2021, Diodato2014}, inequality~\cite{Moretti2012}, and firm success~\cite{Jara-Figueroa2018}. In order to effectively navigate sustainable development and net-zero transitions, policy makers must have a thorough understanding of the spatial distribution of potential labour market implications~\cite{Lim2023}. Rather than looking at skill and spatial frictions in isolation, policy makers should take them into account simultaneously~\cite{Moritz2021, Farinha2019, Vandeplas2022} in order to facilitate the green transition and avoid political pushback for displaced sectors.

Network science can be a powerful tool for understanding these labour mobility frictions. At the industry level, Neffke, Otto, and Weyh~\cite{Neffke2017} find the network of inter-industry labour flows is sparse, stable over time, and predictive of industry growth. O'Clery and Kinsella~\cite{oclery2022modular} find that clustered industries require similar skills, and increased worker availability within an industry cluster is a predictor of employment growth. Occupational mobility networks concisely represent possible transitions between occupations. Cheng and Park~\cite{cheng2020flows} show that communities in such networks are often groups of occupations with similar skills. Toubøl and Larsen~\cite{Toubol2017} show that these communities also group together occupations with similar demographics such as wages and gender make-up. Networks of skills and work activities have also been used to study polarisation in the labour market and job transitions~\cite{alabdulkareem2018unpacking, Aufiero2024,Mealy2018}.

Bucker et al.~\cite{bucker2023employment} used network analysis to show that there can be skill mismatches in a rapid decarbonisation of the US power sectors. They stressed the importance of developing methods that can account for the temporal dynamics in the demand and frictions faced by certain jobs, as some occupations see only a short-term boost in demand, followed by a decline. To this end, Fair and Guerrero~\cite{Fair2023} developed an agent-based model that is able to generate a network of occupational, sectoral, and regional frictions which they show can be used to study future demand shocks in the UK without relying on historic job transitions. Moro et al.~\cite{MoroEsteban2021Urpi} develop another agent-based labour market model with occupations connected using a network where edges are based on the similarity of skills. Their work shows that, during a demand shock, workers in a city with more connections to their occupation are able to find employment more easily than workers in a less connected city. Our work builds upon this literature. We incorporate networks and dynamics to study regional and occupational frictions at a high level of granularity, and take into account non-linear changes in demand for jobs over time. We apply a novel agent-based model of the labour market that explicitly matches individual workers to jobs. 

%3.
%% METHODS
Our analysis extends the occupational labour market model developed by del Rio-Chanona et al.~\cite{delRioChanona2021} to include geographical mobility. We follow Mealy et al.~\cite{Mealy2018} in using historical labour market data to estimate labour market frictions and account for worker transitions between both detailed occupations and regions for the first time.\footnote{
To avoid confusion with the five macroregions of Brazil, we use region here to refer to the 16 regions, defined by Ferreira Filho and Hanusch~\cite{Hanusch2022}. The regions are either states, for example Bahia, or groups of states, for example RSudeste, which is made up of Minas Gerais, Esp\'irito Santo, and Rio de Janeiro. 
} We apply our framework to labour demand scenarios taken from two growth pathways for Brazil with differing environmental impacts. The scenarios model increased productivity in agriculture (\textit{Agriculture growth path}), which leads to higher emissions and slightly reduced deforestation, and manufacturing (\textit{Manufacturing growth path}), which leads to lower emissions and a greater reduction in deforestation, compared to a baseline scenario without the productivity increase.

The labour market model simulates the labour market responding to changes in occupation and region level demand. Workers are represented by agents. According to the labour demand scenarios, labour demand is relocated to different occupations and regions. Workers are fired and vacancies are opened in response to the changing labour demand, and unemployed workers apply for a job with probabilities according to the network of historic labour transitions. 

%4.
%% RESULTS

A network of historical transitions between occupations and regions reveals a strong regional structure. We find that the Manufacturing growth path has lower aggregate unemployment than the Agriculture growth path as well as smaller impacts across occupations and regions. Our analysis identifies the skill and spatial mismatches of worker availability and demand during the two growth pathways.

Our results indicate that the lowest-wage occupations are impacted most negatively. Unless the growth path will come with additional help for workers to transition to skills that are in demand, this might lead to growing inequality. Specifically, if the productivity increase in Manufacturing is realised according to our model, retraining for agriculture and repair/maintenance workers will be needed. Targeted policies may be required for workers in these occupations, as well as services, if the Agriculture growth path is followed.

Our results also identifies the locations where these retraining policies should be prioritised, and where worker relocation or local development policies need to be considered to enable smoother sustainable development in Brazil. In order to follow the Manufacturing growth path successfully, regional policies in Mato Grosso, RSul, and the northern regions may need to be considered. If the Agriculture growth path is followed, policies need to be aimed at inland, agriculture focused regions such as Roraima and Rond\^{o}nia, especially as agricultural productivity gains in these two regions cause more deforestation.

\section{Results} \label{sec:results}
Our analysis explores the potential labour market impacts of the Manufacturing and Agriculture growth paths in Brazil. We show that both pathways have increased unemployment and unfilled vacancy rates compared to the baseline, with the Agriculture growth path facing greater reallocation and higher aggregate unemployment. We investigate how these impacts vary across the wage distribution and across regions and occupations.

%%%%%%
\subsection*{A network perspective on skill and spatial labour frictions}

\begin{figure}[h!]
     \centering
     \begin{subfigure}[b]{0.45\textwidth}
         \centering
         \includegraphics[width=\textwidth]{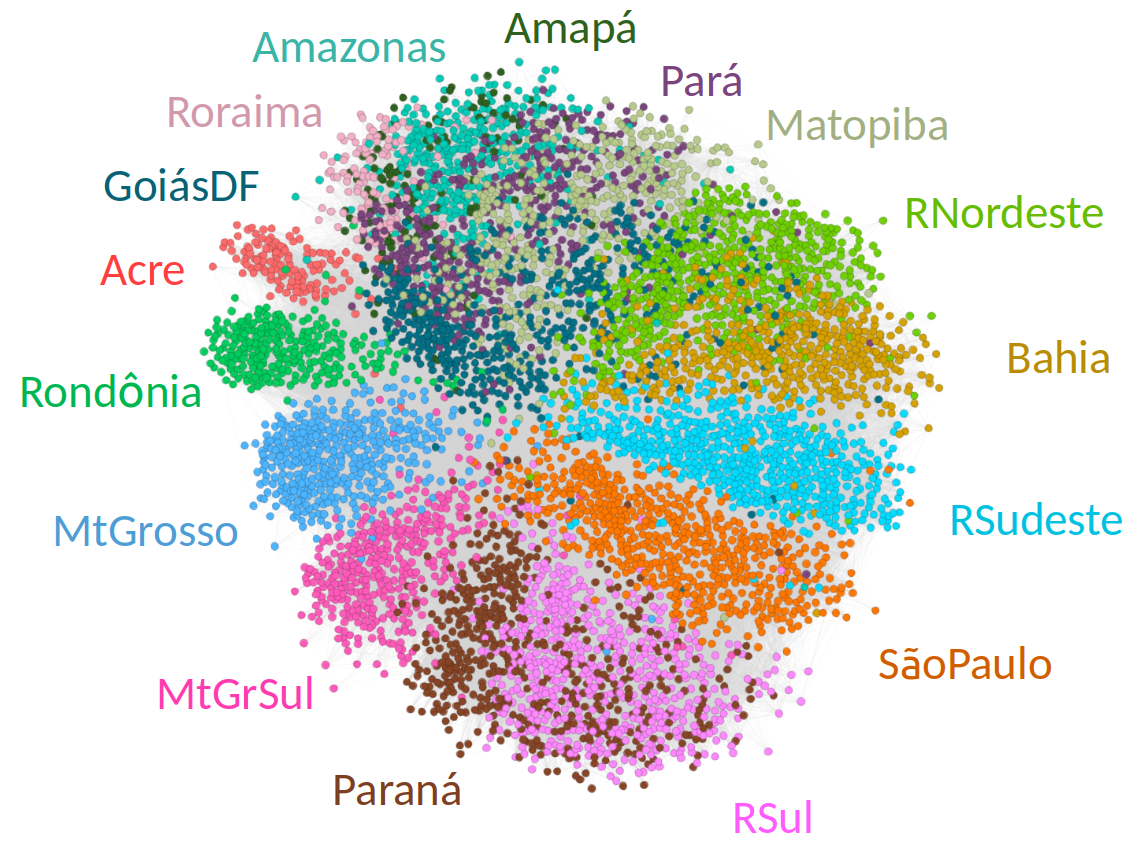}
         \caption{Regional occupational mobility network, coloured by region}
         \label{fig:omng_region}
     \end{subfigure}
     \hfill
     \begin{subfigure}[b]{0.45\textwidth}
         \centering
         \includegraphics[width=\textwidth]{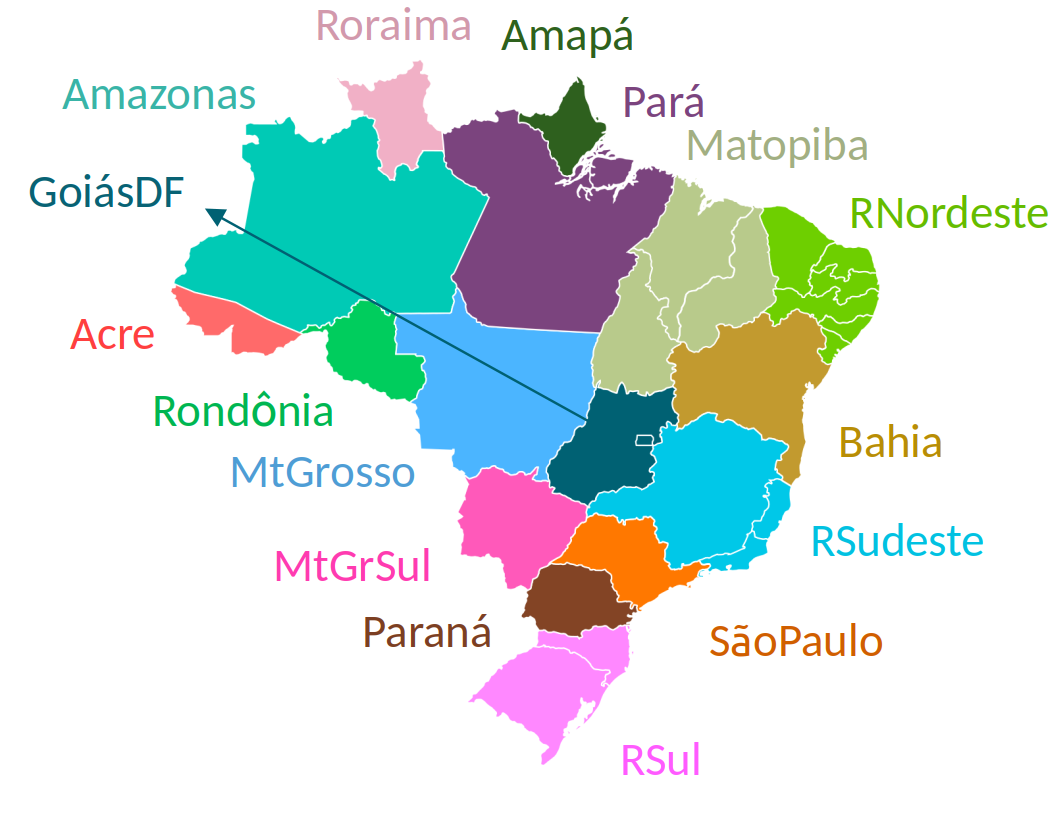}
         \caption{Brazil's states coloured by region}
         \label{fig:omng_map}
     \end{subfigure}

     \begin{subfigure}[b]{0.78\textwidth}
         \centering
         \includegraphics[width=\textwidth]{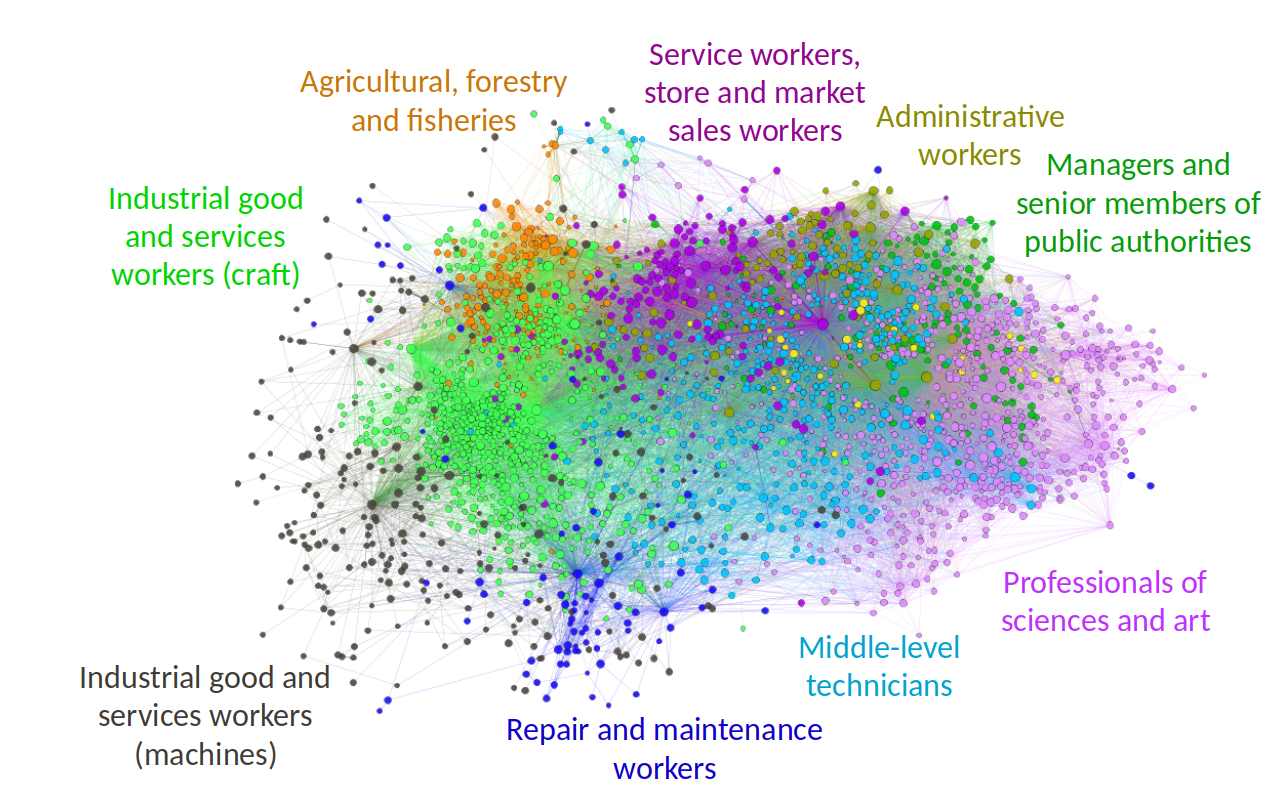}
         \caption{National occupational mobility network, coloured by occupation group}
         \label{fig:omng_occ}
     \end{subfigure}
    \caption{\textbf{Regional occupational mobility network for Brazil.} %Nodes represent occupations and edges represent transitions made between 2011--2019. The node size corresponds to employment in 2018. 
    In (a) the full regional occupational mobility network, in (b) a map of Brazil with states coloured similarly to (a), and in (c) the national occupational mobility network, without disaggregation by state.
    }
    \label{fig:omng}
\end{figure}

We find that both regional and occupational mobility strongly influence labour flows (Figure~\ref{fig:omng}). The regional occupational mobility network (Figure~\ref{fig:omng_region}), which captures the historical transitions workers have made between occupation-region pairs, strongly resembles the geographical map of Brazil (Figure~\ref{fig:omng_map}). This signals inter-regional links are driven by geographical factors, such as moving costs and familial ties.\footnote{
We use the Force Atlas 2~\cite{Jacomy2014} algorithm to position the nodes in these visualisations. Nodes repel each other like charged particles while an edge acts as an attraction force. The algorithm only uses the network structure with no additional information about the nodes.
} 
Furthermore, when considering occupation links only (Figure~\ref{fig:omng_occ}), we find that nodes of the same broad category appear close together. This suggests that transitions to other occupational categories are rare and likely more difficult.

We validate these observations by measuring the assortativity of the network across occupations and regions. The assortativity tells us how connected occupation-region pairs are to other pairs with the same occupation or region (see Methods). In the occupation-region network, we find that regional assortativity is 0.77, and the 1-digit occupation assortativity is 0.56, confirming that both regions and occupations influence labour mobility. The larger value for regional assortativity indicates that transitions happen mostly within, rather than between, regions.

\subsection*{Inequality of unemployment outcomes}

\begin{figure}
\centering
    \begin{subfigure}{0.49\textwidth}
        \centering
        \includegraphics[width=0.9\linewidth]
        {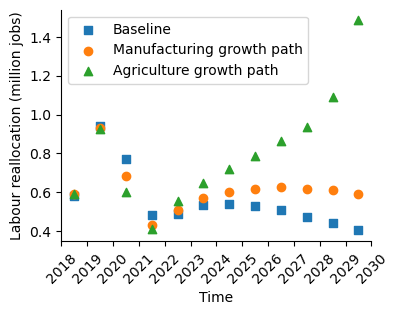}
        \caption{Yearly labour reallocation} 
        \label{fig:labour_reall}    
    \end{subfigure}
    \begin{subfigure}{0.49\textwidth}
        \centering
        \includegraphics[width=0.9\linewidth]
        {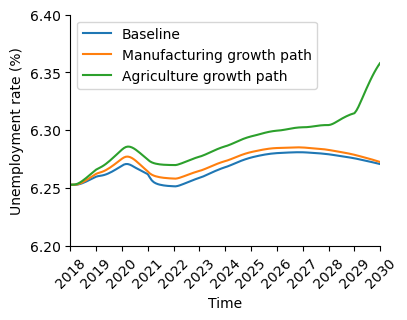}
        \caption{Aggregate unemployment rate} 
        \label{fig:agg_u_omng}    
    \end{subfigure}
    \caption{\textbf{Aggregate reallocation and unemployment rate.} In (a) the volume of labour reallocation per year, and in (b) the aggregate unemployment rate between 2018 and 2030.}
    \label{fig:agg_u_omng_with_labour_reall}    
\end{figure}

Our agent-based model computes how workers move on the full regional occupational mobility network given the labour demand changes induced by the growth pathways. The Agriculture growth path has the greatest labour reallocation (Figure~\ref{fig:labour_reall}), over to 1.4 million new jobs by 2030. The aggregate impacts to unemployment outcomes are relatively small in both scenarios, but the distributional outcomes in the Manufacturing growth path are better (Figure~\ref{fig:agg_u_omng}).

\begin{figure}
\begin{subfigure}{\textwidth}
    \centering
        \includegraphics[width=0.8\textwidth]{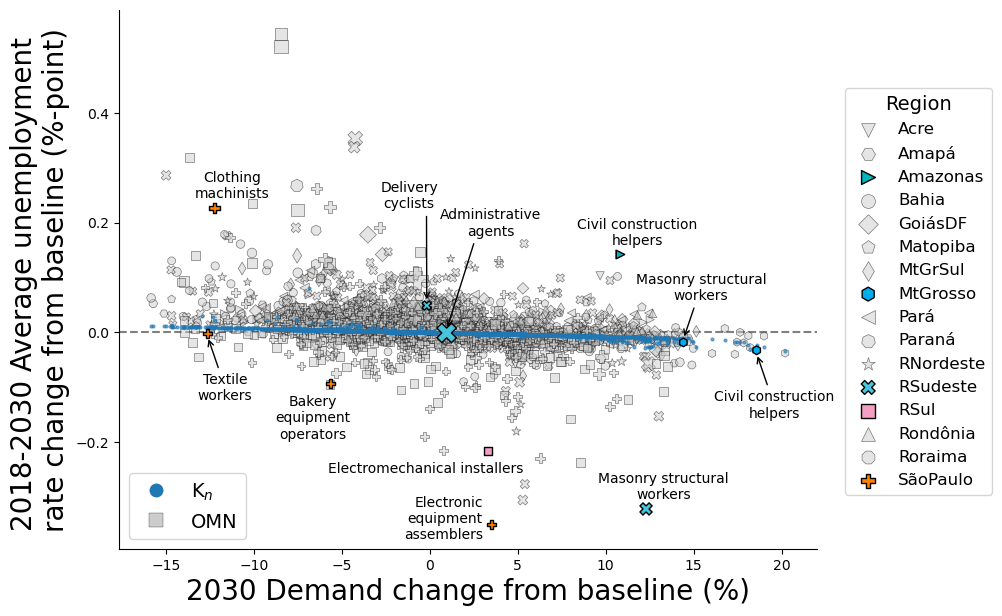}
    \caption{} 
    \label{fig:unemp_manuf_scatter} 
    \end{subfigure}
    ~
    \begin{subfigure}{0.64\textwidth}
    \centering
    \includegraphics[width=\textwidth]{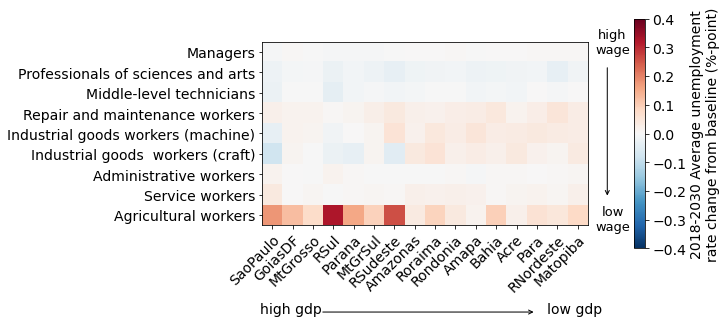}
    \caption{}
    \label{fig:unemp_manuf_heatmap} 
    \end{subfigure}
    ~
    \begin{subfigure}{0.34\textwidth}
    \centering
    \includegraphics[width=\textwidth]
    {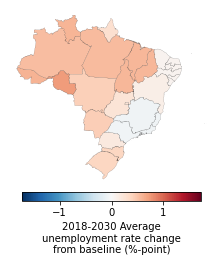}
    \caption{}
    \label{fig:unemp_manuf_map} 
    \end{subfigure}
    \caption{\textbf{Manufacturing growth path unemployment outcomes.} In a) the percentage-point change of the 2018--2030 average unemployment rate from the baseline against percentage demand change compared to baseline in 2030, each data point represents an occupation-region pair, shaped by region and sized proportional to employment in 2018; in b) the percentage-point change of the 2018--2030 unemployment rate from the baseline for each 1-digit occupation and region; and in c) the regional percentage-point change of the average unemployment rate from the baseline.}
    \label{fig:unemp_manuf} 
\end{figure}

\begin{figure}
\begin{subfigure}{\textwidth}
    \centering
        \includegraphics[width=0.8\textwidth]
        {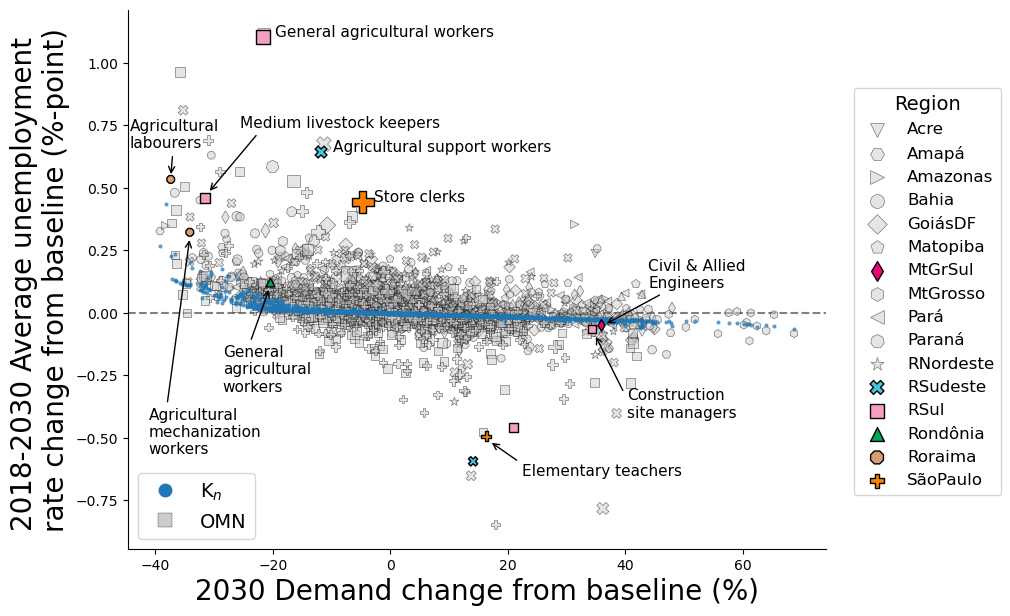}
    \caption{} 
    \label{fig:unemp_agr_scatter} 
    \end{subfigure}
    ~
    \begin{subfigure}{0.64\textwidth}
    \centering
    \includegraphics[width=\textwidth]
    {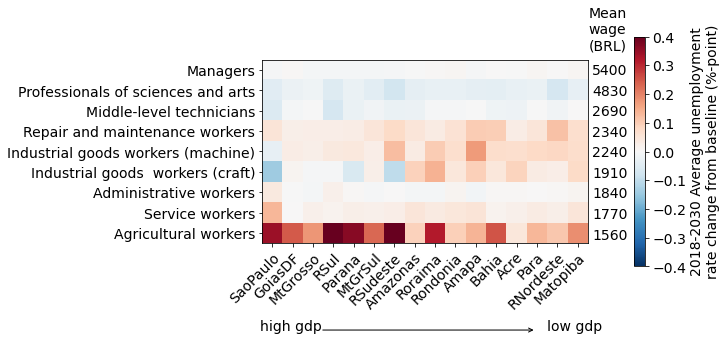}
    \caption{}
    \label{fig:unemp_agr_heatmap} 
    \end{subfigure}
    ~
    \begin{subfigure}{0.34\textwidth}
    \centering
    \includegraphics[width=\textwidth]
    {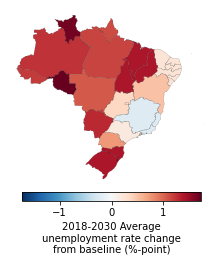}
    \caption{}
    \label{fig:unemp_agr_map} 
    \end{subfigure}
    \caption{\textbf{Agriculture growth path unemployment outcomes.} In a) the percentage-point change of the 2018--2030 average unemployment rate from the baseline against percentage demand change compared to baseline in 2030, each data point represents an occupation-region pair, shaped by region and sized proportional to employment in 2018; in b) the percentage-point change of the 2018--2030 unemployment rate from the baseline for each 1-digit occupation and region; and in c) the regional percentage-point change of the average unemployment rate from the baseline.}
    \label{fig:unemp_agr} 
\end{figure}

There is an inverse relationship between the change in the unemployment rate by occupation-region in relation to the change in labour demand compared to the baseline, as we might expect (Figures~\ref{fig:unemp_manuf_scatter} and~\ref{fig:unemp_agr_scatter}). In both scenarios, about half of the occupation-region pairs face a decrease in labour demand by 2030 compared to the baseline.\footnote{
55.1\% in the Manufacturing growth path and 54.8\% in the Agriculture growth path.
} Our model shows that this translates to 53\% of occupation-region pairs being negatively affected by the Manufacturing growth path (i.e.\ face higher unemployment rates than in the baseline) versus 59\% in the Agriculture growth path.\footnote{
These occupation-region pairs represent 60\% and 56\% of workers, respectively.
}

Occupations with the same change in demand face different unemployment outcomes due to their position in the network and the demand profile of their surrounding occupations. For example, in the Manufacturing growth path (Figure~\ref{fig:unemp_manuf_scatter}), `Textile workers' and `Clothing machinists' in S\~ao Paulo face a similar demand change but different unemployment effects -- the former is almost unaffected by the scenario, while the latter is negatively affected. `Textile workers' in S\~ao Paulo are able to mitigate the decrease in demand for workers in their occupation because they are able to find employment elsewhere. In contrast, `Clothing machinists' in S\~ao Paulo are unable to find employment elsewhere, as many of their closest other options also have a decrease in demand. These complex network effects constitute second order effects that introduce variance in unemployment outcomes. 

To highlight these effects, we show the results for the case with no worker frictions in blue circles in Figures~\ref{fig:unemp_manuf_scatter} and~\ref{fig:unemp_agr_scatter}. Here, we run the model with every occupation-region pair connected to all other occupation-regions with equal weights. Unemployed workers are able to find a new job in any occupation-region pair. In this case, there is little variation for occupation-regions with the same change in demand and little impact on the unemployment rate compared to the baseline. This is because unemployed workers do not face any skill or spatial frictions and so can apply for a job in any occupation-region pair, greatly reducing the friction within the model.

`Store clerks' in S\~ao Paulo have relatively low mobility and their closest options in alternative occupations and regions also experience a decrease in demand. Therefore, in the Agriculture growth path (Figure~\ref{fig:unemp_agr_scatter}), workers in this occupation-region see a significant increase in the unemployment rate compared to the baseline for a relatively small decrease in overall demand. We find that `store clerks' in S\~ao Paulo have few options to react to this small decrease. We find similar patterns for `Agricultural support workers' and other agricultural workers, suggesting that the productivity gains in the agriculture sector may not translate into better outcomes for agriculture workers.

\subsubsection*{Skill mismatches}
High-wage occupations are better positioned to benefit from both scenarios. We find that the best outcomes tend to be concentrated in occupations with higher wages in both growth paths (Figures~\ref{fig:unemp_manuf_heatmap} and~\ref{fig:unemp_agr_heatmap}). This can increase inequality in labour market outcomes and, subsequently, income and welfare.  In the Manufacturing growth path, `Middle-level technicians’, `Professionals of sciences and arts’, and `Industrial goods (craft) workers’ are the occupational groups most likely to benefit while occupational groups with lower wages face increased unemployment rates. In the Agriculture growth path, the high-wage occupational groups, `Managers', `Professionals of sciences and arts', and `Middle-level technicians', face lower unemployment than in the baseline while `Agriculture, forestry and fisheries workers', the occupational group with the lowest mean wage, face a large increase in the unemployment rate, due to their position in the network and decreasing demand (despite the productivity increase). This disparity in unemployment outcomes across the wage distribution is caused by low mobility and by being surrounded by other decreasing-demand occupations.

\subsubsection*{Spatial mismatches}
At the region level, we find that most regions see an increase in the aggregate unemployment rate compared to the baseline (Figures~\ref{fig:unemp_manuf_map} and~\ref{fig:unemp_agr_map}). In the Manufacturing growth path, we find an increase in the unemployment rate in all regions except S\~ao Paulo and RSudeste. In the Agriculture growth path, the unemployment rate increases in all regions except RSudeste. We can also see that better unemployment outcomes are focused in the coastal and metropolitan regions, while the more rural, Amazonian regions face the largest increases in unemployment. In the Agriculture growth path, Roraima and Rond\^{o}nia are particularly badly affected, with the average unemployment rate increasing by 1.66 and 1.72 percentage-points compared to the baseline.

\subsubsection*{Contribution of skill versus spatial}
Unemployment rate changes vary much more by occupation than by region. Most broad occupation groups have similar unemployment outcomes across all regions (Figures~\ref{fig:unemp_manuf_heatmap} and~\ref{fig:unemp_agr_heatmap}). Decomposing the total variance of unemployment outcomes, we find that differences between regions explain little of the heterogeneity in outcomes.\footnote{
It is worth noting that this result is consistent with the fact that the productivity shocks simulated in each scenario are homogeneous across regions and that the between-region components of the demand shocks are relatively low as well.
} Indeed, occupation network effects, as captured by the between-occupation component of the variance, explain about 43\% and 46\% of the variance in unemployment rates in the Manufacturing and Agriculture growth paths respectively (Figure~\ref{fig:variance_decomp_percent_anna_temp}).

\subsection*{Unfilled job vacancies risk slowing down development}

\begin{figure}
\begin{subfigure}{\textwidth}
    \centering
        \includegraphics[width=0.8\textwidth]{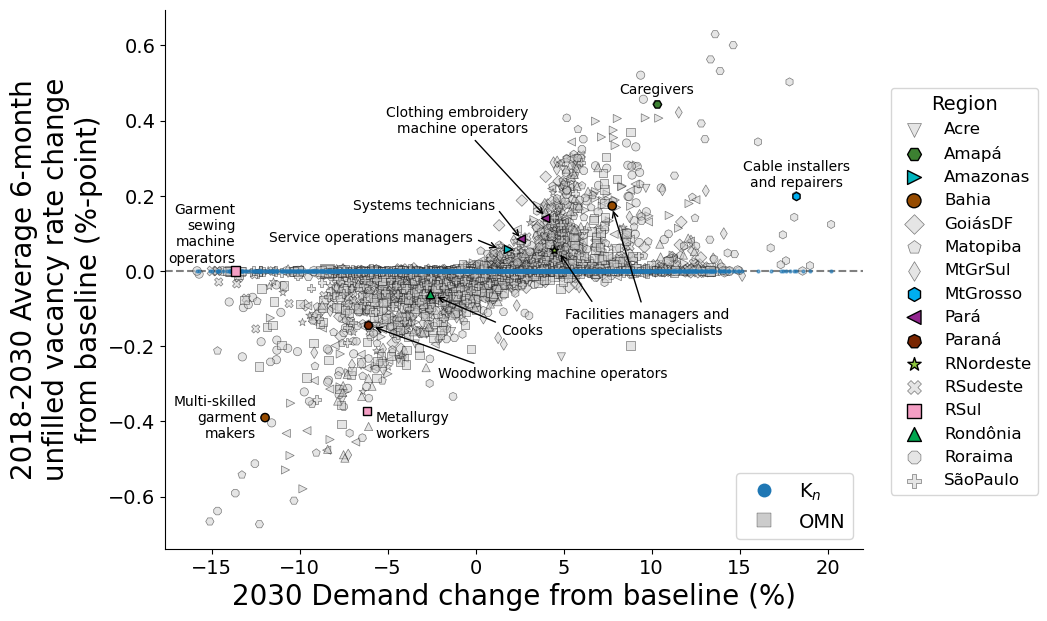}
    \caption{} 
    \label{fig:vacs_manuf_scatter} 
    \end{subfigure}
    ~
    \begin{subfigure}{0.64\textwidth}
    \centering
    \includegraphics[width=\textwidth]{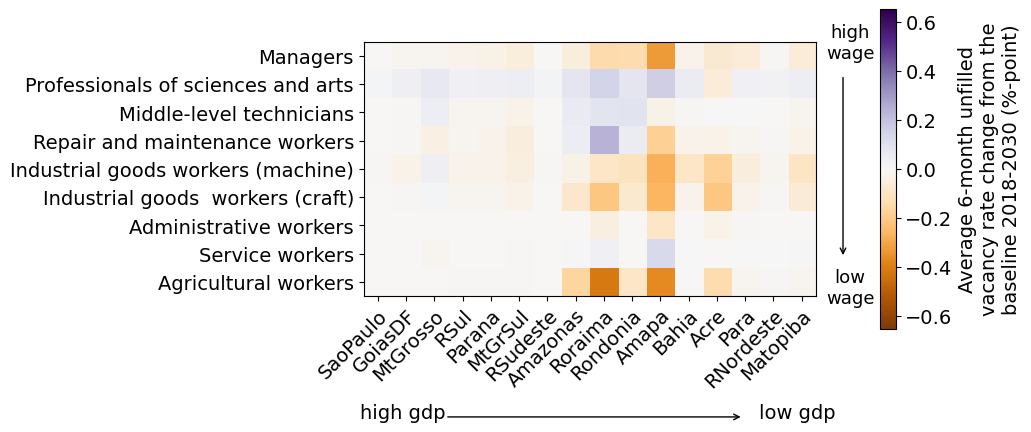}
    \caption{}
    \label{fig:vacs_manuf_heatmap} 
    \end{subfigure}
    ~
    \begin{subfigure}{0.34\textwidth}
    \centering
    \includegraphics[width=\textwidth]{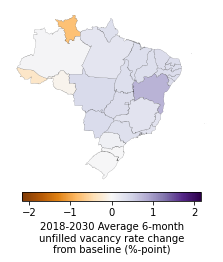}
    \caption{}
    \label{fig:vacs_manuf_map} 
    \end{subfigure}
    \caption{\textbf{Manufacturing growth path vacancy outcomes.} In a) the percentage-point change of the 2018--2030 average unfilled vacancy rate from the baseline against percentage demand change compared to baseline in 2030, each data point represents an occupation-region pair, shaped by region and sized proportional to employment in 2018; in b) the percentage-point change of the 2018--2030 unfilled vacancy rate from the baseline for each 1-digit occupation and region; and in c) the regional percentage-point change of the average unfilled vacancy rate from baseline.}
    \label{fig:vacs_manuf} 
\end{figure}

\begin{figure}
\begin{subfigure}{\textwidth}
    \centering
        \includegraphics[width=0.8\textwidth]
        {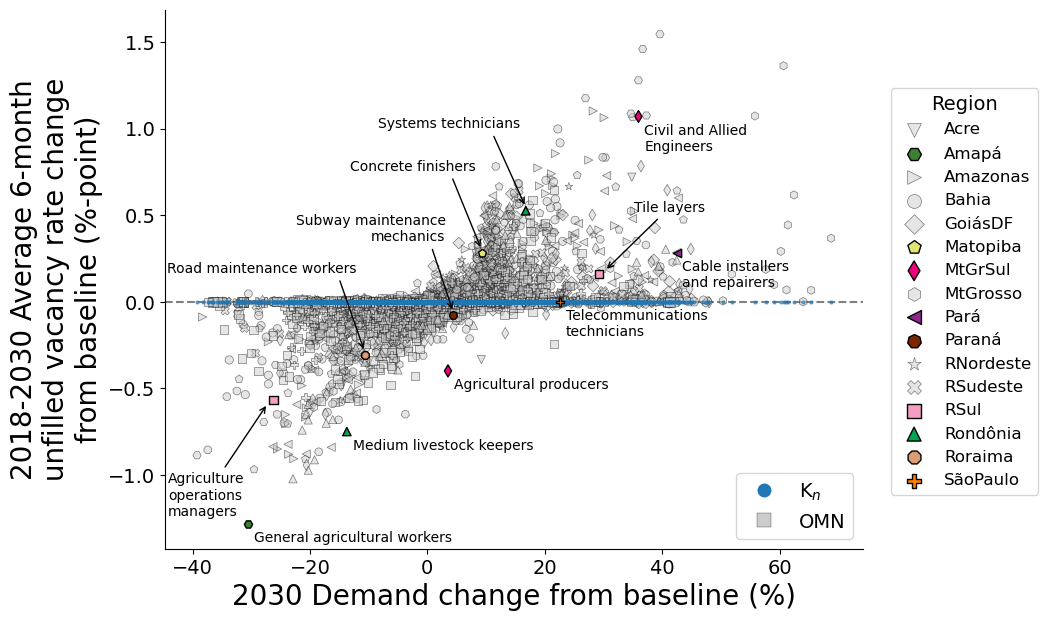}
    \caption{} 
    \label{fig:vacs_agr_scatter} 
    \end{subfigure}
    ~
    \begin{subfigure}{0.64\textwidth}
    \centering
    \includegraphics[width=\textwidth]
    {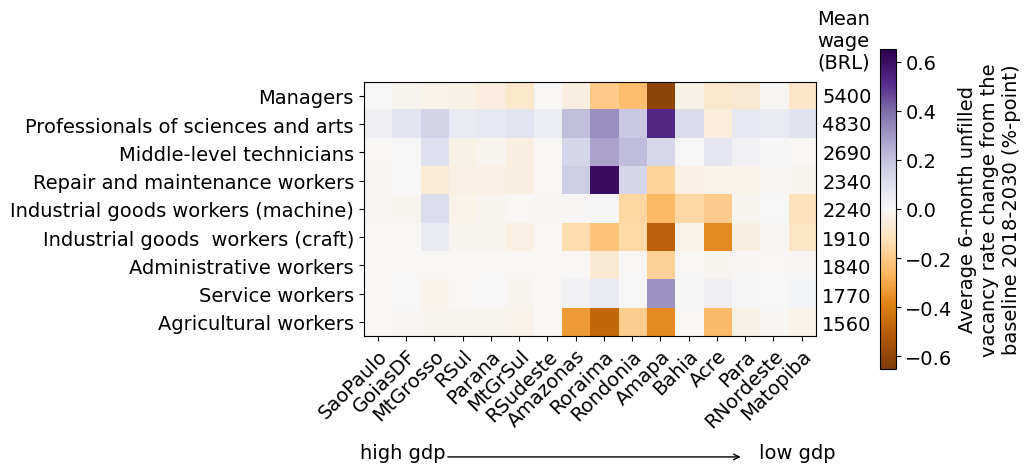}
    \caption{}
    \label{fig:vacs_agr_heatmap} 
    \end{subfigure}
    ~
    \begin{subfigure}{0.34\textwidth}
    \centering
    \includegraphics[width=\textwidth]
    {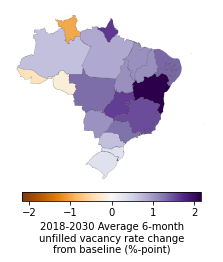}
    \caption{}
    \label{fig:vacs_agr_map} 
    \end{subfigure}
    \caption{\textbf{Agriculture growth path vacancy outcomes.} In a) the percentage-point change of the 2018--2030 average unfilled vacancy rate from the baseline against percentage demand change compared to baseline in 2030, each data point represents an occupation-region pair, shaped by region and sized proportional to employment in 2018; in b) the percentage-point change of the 2018--2030 unfilled vacancy rate from the baseline for each 1-digit occupation and region; and in c) the regional percentage-point change of the average unfilled vacancy rate from baseline.}
    \label{fig:vacs_agr} 
\end{figure}

Higher unemployment rates affect workers whereas an abundance of unfilled vacancies affects employers and risks slowing down the pace of development. In both scenarios there is a positive relationship between changes in labour demand and unfilled vacancy rates (Figures~\ref{fig:vacs_manuf_scatter} and~\ref{fig:vacs_agr_scatter}). In the Manufacturing growth path
we see evidence of friction, with 32\% of occupation-region pairs facing higher unfilled vacancy rates than in the baseline, versus 39\% in the Agriculture growth path.\footnote{
These occupation-region pairs represent 23\% and 20\% of workers respectively.
} 

We also observe second order network effects in the vacancy rate changes for occupation-region pairs with similar demand changes. For example, in the Agriculture growth path, vacancies for `Systems technicians' in Rond\^{o}nia remain unfilled because there are few occupation-region pairs connected to this pair in the network where workers could apply from. Whereas `Telecommunications technicians' in S\~ao Paulo have a strong probability of staying in this occupation-region, which results in open vacancies being filled faster, keeping the unfilled vacancy rate low.

\subsubsection*{Skill mismatches}
Unfilled vacancy rates are concentrated in professional and technical occupations (Figures~\ref{fig:vacs_manuf_heatmap} and~\ref{fig:vacs_agr_heatmap}). Typically, if the unemployment rate increased during the growth path for an occupation group, the unfilled vacancy rate is lower (and vice versa). This relationship aligns with expectations: if there were more job openings for unemployed workers to apply to or fill, the unemployment rate could decrease. For the Manufacturing growth path, we do find vacancies that are hard to fill in high wage occupations, where the unemployment rates were lower than the baseline. However, management occupations face increased unfilled vacancies but saw little impact to their unemployment rate. This indicates that workers in management occupations are able to cope but employers of these highly skilled workers are not. Additionally, service occupations have both increased unemployment and increased unfilled vacancy rates meaning workers and employers of services could be negatively impacted by the Manufacturing growth path. (See Table~\ref{fig:table_results} for aggregate occupation-level outcomes.)

In the Agriculture growth path, firms hiring in the occupational groups with higher wages are likely to experience more friction and unfilled vacancies than in the baseline. It is noteworthy that in the scenario driven by productivity gains in agriculture, the vacancy rates for agricultural workers face the largest decrease compared to the baseline, in line with the higher levels of unemployment likely to be experienced by this group.

\subsubsection*{Spatial mismatches}

At the region level, we find that most regions experience an increase in the unfilled vacancy rate in both scenarios compared to the baseline (Figures~\ref{fig:vacs_manuf_map} and~\ref{fig:vacs_agr_map}). Roraima, Acre, and Rond\^{o}nia see a decreased rate of unfilled vacancies compared to the baseline in both scenarios, while all other regions have increased rates of unfilled vacancies. Complementary to the region-level unemployment rates, the largest increase in unfilled vacancies is seen in the coastal regions, where unemployment outcomes are better.

\section{Discussion} \label{sec:discussion}

Understanding the labour market impacts of different development pathways and policies, alongside their environmental impacts, is vital for policy development and public acceptance, and more focus on developing countries' policies is needed to mitigate global climate change~\cite{Caucheteux2025}. The unemployment and vacancy impacts of development policies depend on labour mobility frictions. We introduced a regional occupational mobility network for Brazil which accounts for frictions when changing occupation and/or region. We linked this network to an external macro-economic model and a labour market agent-based model to investigate the labour market impacts of two development pathways in Brazil -- one in which total factor productivity increases in the manufacturing sector and another with the productivity increase in the agriculture sector. 

We identified occupation-region pairs that may require the attention of policy makers due to increased unemployment or rising unfilled vacancy rates. Occupation-region pairs with high unemployment are a concern due to the social and economic damage that unemployment can cause. Meanwhile, addressing vacancy rates is important to support development progress and ensure that the right workers with the right skills are available to fulfil demand. 

We visualise the skill and spatial frictions within the labour market of Brazil in a novel network approach. We find a rich regional structure within the network, with inter-region transitions tightly reflecting the geography of Brazil. This network enables us to show and model key empirical labour market structures that are not captured by standard modelling approaches. The contribution of the network to understanding the impact of different growth paths is highlighted when we compare the results of our model with and without labour frictions. We identify occupation-region pairs at risk of negative impacts due to the interaction of the scenarios and their position in the regional occupational mobility network.

Our results suggest that both pathways present some risk to workers during the transition period, with better aggregate effects seen in the Manufacturing pathway compared to the baseline than in the Agriculture pathway. The pathways also come with challenges for employers, who may struggle to find workers with the appropriate skills in the required region. Therefore, training and other local development policies may be important to support sustained growth. 

%region paragraph
Contributing to these challenges is the difference between the regions with increased unemployment and those with increased unfilled vacancies. The regions with increased unemployment are coastal, and have historically been more attractive to workers than the interior. Carvalho and Inácio de Moraes~\cite{Carvalho2021} find the Brazilian Coastal and Marine economies represented an outsized 19.0\% of Brazil GDP in 2015, with much of this activity dominated by services. The regional disconnect between increased unemployment and increased unfilled vacancy rates is in line with Lim et al.~\cite{Lim2023} who find that, in the US, the new green jobs being created are not necessarily in the same location as the brown jobs. As distributional disparities such as these could fuel discontent~\cite{rodriguez2023green}, attending to these regional inequalities could be crucial to the political sustainability of any transition path. 

%skill/wage paragraph
Another possible inequality that could be exacerbated by these growth pathways in across the wage distribution. We find that workers at the bottom end of the wage distribution could be most adversely affected and therefore may require assistance from policy makers if these growth pathways are realised. This possible inequality of sustainable development pathways has been identified before~\cite{sato2023skills}. Our results are further evidence that retraining policies are likely to be needed to mitigate the negative impact of transition scenarios on low-wage workers~\cite{Vandeplas2022}.

This work has presented one of the most granular labour market models available but leaves plenty of room for future work. The method could be extended by more tightly coupling the labour market agent-based model with an external macro-economic model, such as \cite{dosi2020labour,wiese2024forecasting}, to allow for feedback between the industrial output and labour market frictions. This would create the possibility of quantifying a slowdown in growth relating to unfilled vacancies. Additionally, the worker mobility data we use covers only the formal labour market in Brazil, about 67\%~\cite{Ulyssea2018} of the total labour force. Although the dataset we use accurately reflects nationwide demographics such as gender and age~\cite{DeNegri2001}, the lack of informal transitions might be particularly consequential for occupations with high levels of informality (such as in the Services, Agricultural, and Industrial craft occupational groups) which are under-represented in our data. Additional data, such as from the Brazil National Household Sample Survey (PNAD), could be used to understand and correct the sample bias in the RAIS data~\cite{Rivera2013}.

Nonetheless, our method enables us to identify the labour market outcomes across regions and occupations with high granularity. Our results for Brazil echo those of Bergant, Mano, and Shibata~\cite{Bergant2022} for the US: a green transition may not dramatically shift employment geographically but requires different skills. We contribute a new model that provides important insights that policymakers need to account for when thinking about future developmental pathways; in particular, how workers at the bottom end of the wage distribution could be most adversely affected and how skill shortages at the top end of the wage distribution could slow down the pace of progress. We find that skills development programs and policies should therefore be put in place sooner rather than later.

\section{Methods} \label{sec:method}

\subsection{Regional occupational mobility network} \label{subsec:omn}

We extend the work of Mealy, del Rio-Chanona, and Farmer~\cite{Mealy2018} on \textit{occupational mobility networks} by incorporating regional mobility. Each node in the regional occupational mobility network represents an occupation-region pair. Weighted edges between them are informed by empirically observed job transitions.

We construct the regional occupational mobility network from 2011 to 2019 with 409 occupations and 16 regions using data from Rela\c{c}\~{a}o Anual de Informa\c{c}\~{o}es Sociais (RAIS). These 409 occupations are a hybrid level of the CBO2002 classification. We start with the 4-digit level of the CBO2002 classification, which as 570 occupations. In order to have a fully connected component and a consistent list of occupations across all regions, we merge some of these occupations into hybrid 2- and 3-digit codes. This merging results in 409 unique occupations.\footnote{
We also group the occupations into ten 1-digit occupational groups, such as `Professionals of sciences and arts'. 
} The RAIS data contains all active employer-employee contracts in each year in the formal labour force in Brazil, which accounts for about 67\%~\cite{Ulyssea2018} of the total labour force in Brazil. 

A link between two occupation-region pairs is the empirical probability of moving between them, observed in the data between 2011 to 2019. Let~$T_{i\alpha,j\beta}$ be the number of workers that transitioned from occupation $i$ in region~$\alpha$ to occupation $j$ in region~$\beta$ between 2011 and 2019. The probability a worker transitions from occupation-region~$i\alpha$ to occupation-region~$j\beta$ is then
\begin{align}
    A_{i\alpha,j\beta} = \frac{T_{i\alpha,j\beta}}{\sum_{i,\alpha}{T_{i\alpha,j\beta}}}. \label{eq:omn_adj}
\end{align}

\subsection{Assortativity}
Assortativity measures the tendency of nodes to be connected with other nodes similar in attribute $x$. Assortativity of a network, $A_{i\alpha,j\beta}$ with attribute $x$ was originally defined by Newman~\cite{Newman2003} and later extended to weighted networks by Yuan, Yan, and Zhang~\cite{Yuan2021}. We use weighted categorical assortativity, which is defined by B{\"u}cker et al.~\cite{bucker2023employment} as
\begin{align}
    r = \frac{\sum\limits_{i} (e_{ii} - a_i b_i)}{1-\sum\limits_{i}  a_i b_i} 
\end{align}
where $e_{ij}$ is the fraction of all edge weights that join nodes with value $i$ to $j$, $a_i$ and $b_j$ are the fraction of edges that start at $i$ and end at $j$ respectively.

\subsection{Labour market model} \label{subsec:abm}
We extend the agent-based labour market model developed by del Rio-Chanona et al.~\cite{delRioChanona2021} to include geographical mobility. Using the regional occupational mobility network, we model worker mobility between occupation-region pairs.

Following del Rio-Chanona et al.~\cite{delRioChanona2021}, we track the number of workers employed and unemployed, and the number of vacancies open, at each occupation-region pair. At time~$t$, let~$e_{i\alpha;t}$ be the number of workers employed in occupation-region pair~$i\alpha$, $u_{i\alpha;t}$ the number of workers unemployed in occupation-region~$i\alpha$, and~$v_{i\alpha;t}$ the number of vacancies open in occupation-region~$i\alpha$. An unemployed worker is considered to be in the occupation-region pair in which they were most recently employed. 

Workers quit or lose their job and vacancies are opened by two processes: a spontaneous process and a state-dependent process. In the spontaneous process, workers are separated and vacancies are opened with probability~$\delta_u$ and~$\delta_v$ respectively. In the state-dependent process, workers are separated \textit{or} vacancies are opened depending on the difference between the realised demand,~$d_{i\alpha;t} = e_{i\alpha;t}+v_{i\alpha;t}$, and the target demand, $d^{\dagger}_{i\alpha;t}$. The target demand takes the form of labour demand for each occupation-region pair at each time step and is where the link to the external macro model is introduced. 

The labour flow,~$F_{i\alpha,j\beta;t+1}$, is the number of workers hired in occupation-region~$j\beta$ who were previously unemployed in occupation-region~$i\alpha$. At time~$t$, let~$b_{i\alpha;t}$ be the number of workers separated from occupation-region~$i\alpha$ and let~$c_{i\alpha;t}$ be the number of vacancies that are opened in occupation-region~$i\alpha$. Then the equations governing the flow of workers around the network are
\begin{align}
    e_{i\alpha;t+1} &= e_{i\alpha;t} - \underbrace{b_{i\alpha;t+1}}_{\text{separated workers}} + \underbrace{\sum_{j,\beta} F_{j\beta,i\alpha;t+1},}_{\text{hired workers}} \label{eq:gov_emp} \\
    u_{i\alpha;t+1} &= u_{i\alpha;t} + \underbrace{b_{i\alpha;t+1}}_{\text{separated workers}} - \underbrace{\sum_{j,\beta} F_{i\alpha,j\beta;t+1},}_{\text{transitioning workers}} \label{eq:gov_unemp} \\
    v_{i\alpha;t+1} &= v_{i\alpha;t} + \underbrace{c_{i\alpha;t+1}}_{\text{opened vacancies}} - \underbrace{\sum_{j,\beta} F_{j\beta,i\alpha;t+1}.}_{\text{hired workers}} \label{eq:gov_vac}
\end{align}

\subsection{Transition scenarios} \label{subsec:cge}
We use two scenarios that model growth pathways for Brazil plus a baseline scenario from Ferreira Filho and Hanusch~\cite{Hanusch2022}. They use a computable general equilibrium model that includes a land-use module to look at different scenarios for Brazil from 2018 to 2030. Brazil is among the largest greenhouse gas emitters~\cite{UNEP2021}, primarily due to emissions related to deforestation and agriculture~\cite{SEEG2023}. Brazil aims, as set out in its NDC (Nationally Determined Contribution)~\cite{ndc_adjust2023}, to develop sustainably while lowering emissions by 53\% in 2030 compared to 2005 levels and reach climate neutrality by 2050. Ferreira Filho and Hanusch try to understand how different sectoral productivity pathways relate to these decarbonisation targets.

In this paper, we will look at two of these scenarios: 1) the \textit{Manufacturing growth path} assumes an annual 0.5\% total factor productivity growth in the manufacturing sector across Brazil; 2) the \textit{Agriculture growth path} assumes a 0.5\% total factor productivity increase in the agriculture sector across Brazil. Ferreira Filho and Hanusch's results indicate that the Manufacturing growth path is associated with a 3.9\% increase in GDP, 0.8million hectares less deforestation, and over 67,000kT less CO$_2$ emissions in Brazil compared to the baseline scenario with no change to total factor productivity, representing the status quo. The Agriculture growth path is projected to have an increase of 1.8\% in GDP, 0.3million hectares less deforestation, but 18,000kT more CO$_2$ emissions compared to the baseline scenario. While Ferreira Filho and Hanusch find that agricultural productivity across Brazil generally leads to less deforestation, emissions arise due to the high carbon intensity of agriculture. Deforestation is primarily driven by agricultural productivity in the Amazon states, where the \textit{Jevons effect} operates along the agricultural frontier. In contrast, the \textit{Borlaug effect} holds nationwide, increasing productivity without necessarily causing deforestation. Finally, they identify that for Brazil to achieve higher income levels, it will need to focus on a more balanced structural transformation, rather than relying on agriculture, which has historically been Brazil's competitive strength.

We convert the labour demand output from the two sectoral scenarios and the baseline scenario to target demand in our model. First, we convert sector-level labour demand data to occupation-level labour demand for each scenario, using a breakdown of occupational employment per industry in 2018 and 2019. We assume this breakdown to be stable for the duration of the scenario. The CGE model includes an assumption on population growth, but the labour market agent-based model can only handle a fixed population size. Therefore, we normalise all three scenarios with the total employment in the baseline scenario. If total labour demand for occupation-region pair $i\alpha$ in year $\tau$ in scenario $s$ is $D_{i\alpha,\tau,s}$, then the adjusted target demand for labour, $D^*_{i\alpha,\tau,s}$, is
\begin{align}
    D^*_{i\alpha,\tau,s} = \frac{D_{i\alpha,\tau,s}\cdot \sum_{i,\alpha}D_{i\alpha,2018,baseline}}{\sum_{i,\alpha}D_{i\alpha,\tau,baseline}}. \label{eq:normalisation}
\end{align} 

\begin{figure}
     \centering
     \begin{subfigure}[b]{0.65\textwidth}
         \centering
        \includegraphics[width=\textwidth]{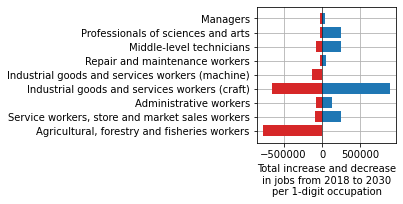}
         \caption{Manufacturing growth path}
         \label{fig:manuf_occ_demand}
     \end{subfigure}
    \begin{subfigure}[b]{0.34\textwidth}
         \centering
         \includegraphics[width=0.88\textwidth]{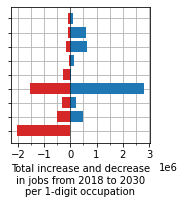}
         \caption{Agriculture growth path}
         \label{fig:agr_occ_demand}
     \end{subfigure}
    \caption{\textbf{Occupation level labour demand change.} Labour demand change per 1-digit occupational group for (a) the Manufacturing growth path, and (b) the Agriculture growth path.}
    \label{fig:labour_demand}
\end{figure}

Within each year, we interpolate the demand per model timestep linearly to calculate target demand ($d^{\dagger}_{i\alpha;t}$). The total number of jobs created and destroyed differs for each occupational group (Figure~\ref{fig:labour_demand}).\footnote{
We count the total jobs created from 2018 to 2030 in the occupation-region pairs within each 1-digit occupational group where demand increases and we count the total jobs destroyed in those occupation-region pairs where demand decreases.
}
We show the difference in magnitude of reallocation between the two growth paths as well as the qualitative similarities. 

\subsection{Rates}
The unemployment rate of occupation-region pair $i\alpha$ represents the workers that lost their job while working in occupation-region pair $i\alpha$ and have subsequently not found new employment. The average unemployment rate over the study period for occupation-region pair $i\alpha$ is defined as
\begin{align}
    u_{i\alpha;average}(T) = \frac{\sum_{t\in T} u_{i\alpha;t}}{\sum_{t\in T} \left( u_{i\alpha;t}+e_{i\alpha;t} \right)}, \label{eq:unemp_rate}
\end{align}
where $T$ is the set of time steps that correspond to the years of transition, 2018 to 2030. 

The average $x$-month vacancy rate $v^{\left(\geq x \right)}_{i\alpha;average}$ is the fraction of vacancies over total realised demand in occupation-region pair $i\alpha$ that have been open for at least $x$ months, on average over the study period: 
\begin{align}
    v^{\left(\geq x\right)}_{i\alpha;average}(T) = \frac{\sum_{t\in T} v^{\left(\geq x\right)}_{i\alpha;t}}{\sum_{t\in T} \left( v_{i\alpha;t}+e_{i\alpha;t} \right)}. \label{eq:vac_rate}
\end{align}

Throughout this study, we use the average 6-month vacancy rate, but the results are qualitatively unaffected by choosing the duration of an unfilled vacancy to be three, six, or twelve months.

\subsection{Occupation-related versus region-related network effects}
We use variance decomposition to understand the contribution of between-region versus between-occupation variation to the total variation of unemployment and vacancy rate outcomes. Based on~\cite{Baumgarten2020}, we decompose the variance of the unemployment rate $u$ as
\begin{align} \label{eq:var_decomp}
    \underbrace{\frac{1}{N}\sum_{i,\alpha} (u_{i\alpha}-\bar{u})^2}_{\text{total variance}}  = 
        \underbrace{var(\bar{u}_\alpha)}_{\text{between-region}} +
        \underbrace{var(\bar{u}_i)}_{\text{between-occupation}}+
        \underbrace{var(u_{i\alpha}-\bar{u}_\alpha-\bar{u}_i)}_{\text{residual variance}},
\end{align}
where $N=3,533$ is the total number of occupation-region pairs, $\bar{u}_{\alpha}$ is the mean unemployment rate change within region $\alpha$, $\bar{u}_{i}$ is the mean unemployment rate change within occupation $i$, and $\bar{u}$ is the overall mean unemployment rate change across all occupation-region pairs. The variance decomposition for vacancy rates follows analogously.

% \newpage
\begin{appendices}
\section{}
\renewcommand{\thefigure}{A\arabic{figure}}
\setcounter{figure}{0}

\renewcommand{\thetable}{A\arabic{table}}
\setcounter{table}{0}

\begin{figure}[h]
    \centering
    \includegraphics[width=\textwidth]{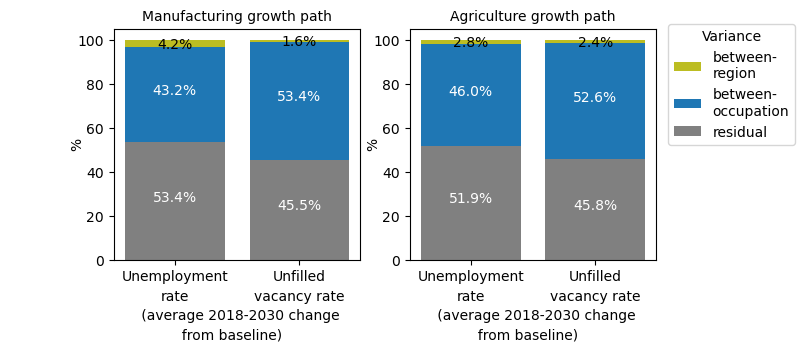} 
    \caption{\textbf{Variance decomposition.} Variance decomposition~\eqref{eq:var_decomp} of the unemployment and 6-month unfilled vacancy rate outcomes for the Agriculture and Manufacturing growth paths.}
    \label{fig:variance_decomp_percent_anna_temp}  
\end{figure}

\begin{table}[h]
    \centering
        \centering
        \includegraphics[width=0.78\textwidth, trim={0cm 0cm 0cm 0cm},clip]{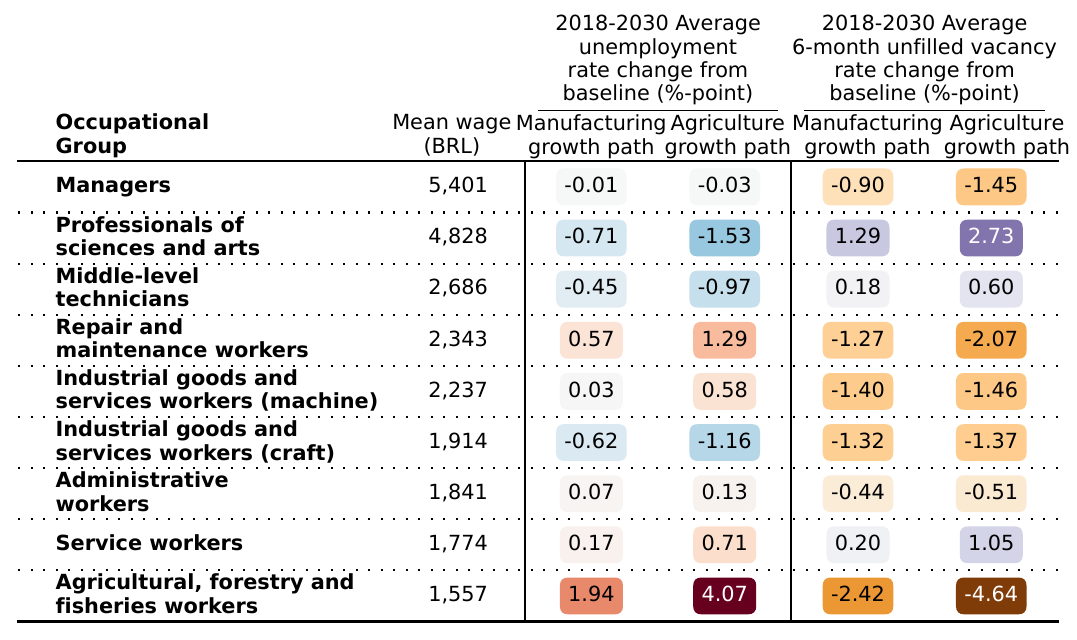} 
        \includegraphics[width=0.1\textwidth, trim={11cm 0cm 0cm 0cm},clip]{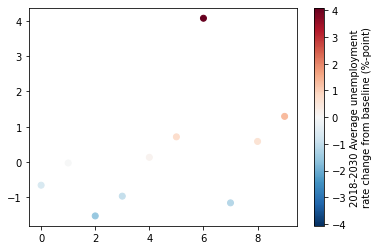}
        \includegraphics[width=0.1\textwidth, trim={11cm 0cm 0cm 0cm},clip]{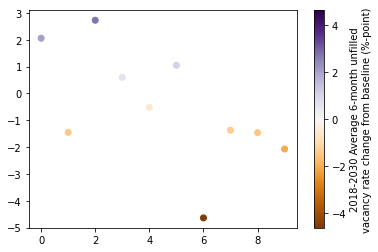}
        \label{fig:occ1table}

    \caption{\textbf{Aggregate outcomes.} Average unemployment and unfilled vacancy rate change compared to the baseline between 2018 and 2030 by occupational group.}
    \label{fig:table_results}
\end{table}

\begin{table}
    \centering
    \begin{tabular}{l|l|l}
         & Agriculture & Manufacturing \\
         \hline
         \makecell[l]{Key growth path \\ characteristics} & \makecell[l]{
          5\% increase in TFP \\ \hspace{3.2mm} in manufacturing sector \\
        % \tabitem  jobs created 2.99m \\ 
        %  \tabitem  jobs lost -2.89m 
         }  & \makecell[l]{
          5\% increase in TFP \\ \hspace{3.2mm} in agriculture sector \\
         % \tabitem  jobs created 1.14m \\
         % \tabitem  jobs lost -1.01m 
         } \\
         \hline
         % Higher unemployment & \makecell[l]{53\% occupation-region pairs \\
         % 56\% workers} & \makecell[l]{59\% occupation-region pairs \\
         % 60\% workers} \\
         % Higher vacancy rates & \makecell[l]{39\% occupation-region pairs \\
         % 23\% workers} & \makecell[l]{32\% occupation-region pairs \\
         % 20\% workers} \\
         % \hline
          \makecell[l]{Top 5 occupation-\\regions with \\ higher unemployment } & \makecell[l]{1. Agricultural support \\ \hspace{3.2mm}workers in RSul \\
         2. Agricultural workers in \\ \hspace{3.2mm}general in RSul \\
         3. Agricultural workers in \\ \hspace{3.2mm}general in RSudeste \\
         4. Agricultural support \\ \hspace{3.2mm} workers in RSudeste \\
         5. Garment sewing machine \\ \hspace{3.2mm} operators in RSul \\
         }
         & 
         \makecell[l]{1. Agricultural support \\ \hspace{3.2mm} workers in RSul \\
         2. Agricultural workers \\ \hspace{3.2mm} in general in RSul \\
         3. Garment sewing machine \\ \hspace{3.2mm} operators in RSul \\
         4. Garment sewing machine \\ \hspace{3.2mm} operators in RSudeste \\
         5. Garment sewing machine \\ \hspace{3.2mm} operators in S\~ao Paulo 
         } 
          \\
         \hline
         \makecell[l]{Top 5 occupation-\\regions with \\ higher vacancy \\ rates } & \makecell[l]{1. Programmers, evaluators, \\ \hspace{3.2mm} and guidance counselors \\ \hspace{3.2mm} in Amapa \\
         2. Teachers in the area of \\ \hspace{3.2mm} pedagogical training in \\ \hspace{3.2mm} higher education in Amapa \\
         3. Elementary school teachers \\ \hspace{3.2mm} in grades five through eight \\ \hspace{3.2mm} in Amapa \\
         4. High school teachers in Amapa \\
         5. Higher education professors of \\ \hspace{3.2mm} architecture and urbanism, engineering, \\ \hspace{3.2mm} geophysics, and geology in Bahia \\
         }
         & 
         \makecell[l]{1. Teachers in the area of \\ \hspace{3.2mm} pedagogical training in \\ \hspace{3.2mm} higher education in Amapa \\
         2. Programmers, evaluators, \\ \hspace{3.2mm} and guidance counselors \\ \hspace{3.2mm} in Amapa \\
         3. Civil and Allied Engineers \\ \hspace{3.2mm} in MtGrosso \\
         4. Elementary school teachers \\ \hspace{3.2mm} in grades five through eight \\ \hspace{3.2mm} in Amapa \\
         5. Child, youth, adult and \\ \hspace{3.2mm} elderly caregivers in Amapa
         }  \\
         
    \end{tabular}
    \caption{\textbf{Summary table.} Top 5 affected occupation-region pairs.} 
    \label{tab:summary_tab}
\end{table}

\end{appendices}

\newpage
\bibliographystyle{abbrv} %plainnat for natbib. plain if not
\bibliography{mybib}

\end{document}